\documentstyle[prl,aps,multicol]{revtex}

\begin{document}
\draft
\title{ Fast Relaxation in a Fragile Liquid under Pressure }

\author{A.~T\"olle$^{1,2}$, H.~Schober$^1$, J.~Wuttke$^3$, 
        O.~G.~Randl$^1$\cite{Randl}, F.~Fujara$^2$}
\address{$^{1}$Institut Laue-Langevin, 38042 Grenoble, France}
\address{$^{2}$Fachbereich Physik, 
               Universit\"at Dortmund, 44221 Dortmund, Germany}
\address{$^{3}$Physik-Department,
               Technische Universit\"at M\"unchen, 85747 Garching, Germany}
\date{\today}
\maketitle

\begin{abstract}
The incoherent dynamic structure factor of ortho-terphenyl has been 
measured by neutron time-of-flight and backscattering technique in the
pressure range from 0.1\,MPa to 240\,MPa 
for temperatures between 301\,K and 335\,K.
Tagged-particle correlations in the compressed liquid decay in two steps.
The $\alpha$-relaxation lineshape is independent of pressure, and
the relaxation time proportional to viscosity.
A kink in the amplitude $f_Q(P)$ reveals the onset of $\beta$ relaxation.
The $\beta$-relaxation regime
can be described by the mode-coupling scaling function;
amplitudes and time scales allow a consistent determination
of the critical pressure $P_c(T)$.
$\alpha$ and $\beta$ relaxation depend in the same way on 
the thermodynamic state;
close to the mode-coupling cross-over,
this dependence can be parametrised by
an effective coupling $\Gamma\propto n T^{-1/4}$.
\end{abstract}

\begin{multicols}{2}


The glass transition can be induced by decreasing the temperature
or by increasing the pressure.
For practical reasons,
most experimental investigations 
concentrate on temperature effects at ambient pressure $P_0=0.1\,$MPa,
although there is a clear interest in studying
supercooled liquid dynamics in the full $P,T$ parameter space.
Variable pressure measurements have been decisive
in demonstrating that density is not the only driving force behind the
glass transition \cite{freeVolume,Leyser_95,Li_95}.
In the microscopic approach of mode-coupling theory \cite{MCT},
both pressure and temperature control the dynamics through variations
of the static structure factor.
Close to the dynamic cross-over, 
the effect of these variations can be expressed 
through a single separation parameter
which however is not given by density alone,
as was confirmed by depolarised light scattering in the fragile liquid cumene
\cite{Li_95}.

Here we report on incoherent neutron scattering 
in the fragile van der Waals liquid ortho-terphenyl 
(OTP, $T_g(P_0)=243\,$K) at variable pressure.
As function of temperature,
the microscopic dynamics of OTP has been studied extensively by
neutron scattering \cite{BartschPetry,WuSqt,ToSqt},
light scattering \cite{Mainz,Cumm}
and molecular dynamics \cite{LeWa,KuWi}.
As function of pressure,
viscosity \cite{Schug},
photon correlation \cite{Fytas_83},
dielectric loss \cite{Naoki_87}, and
specific heat \cite{Leyser_95,Atake_79} data
are available for comparison.
{}From the Debye-Waller factor \cite{BartschPetry,LeWa}
and from direct observation of fast $\beta$ relaxation 
  \cite{WuSqt,ToSqt,Mainz,Cumm}
a cross-over was located at $T_c(P_0)=290\pm\,5$K \cite{eMCT}.


OTP (Aldrich) was purified by repeated 
crystallisation from liquid methanol and vacuum destillation.
An Al7049.T6 high pressure cell of the Institut Laue-Langevin (ILL) 
with a ratio of outer to inner diameter of 2 
allowed us to attain up to 250\,MPa at temperatures up to 335\,K;
helium was used to transmit the pressure.

The experiments were performed using backscattering (IN16) and 
time-of-flight (IN6) spectrometers of the ILL.
Incident wavelengths 6.27\,\AA$^{-1}$ and 5.12\,\AA$^{-1}$
gave resolutions (fwhm) of about 1\,$\mu$eV and 80\,$\mu$eV, respectively.
Vanadium was used to calibrate the detectors 
and to yield the resolution functions.
The measured transmission of the empty pressure cell at $P_0$ was
about 84\%, that of the sample about 88\%.
Self shielding and multiple scattering turned out be negligible 
at large scattering angles.
This was demonstrated by comparison of spectra taken with and without 
the pressure cell at 290\,K:
the normalised intermediate scattering functions $S(Q,t)$ fall onto each
other.
Therefore we base our conclusions exclusively 
on high $Q$ data ($\gtrsim1\,$\AA$^{-1}$).
On IN16 we measured isoviscous and isothermal 
(320\,K) spectra as well as the isothermal (301\,K) elastic intensity.
On IN6 we monitored two different isotherms (316 and 324\,K).
Preliminary results from isochoric measurements have been reported in 
\cite{ToSW97a}.


Before turning to the fast dynamics,
we address the long-time limit of $\alpha$ relaxation;
results on $\alpha$ relaxation will be needed as
input in the analysis of $\beta$ relaxation.

In a first survey, backscattering showed
that the mean relaxation time follows
the pressure dependence of viscosity.
Then, the shape of $\alpha$-relaxation spectra was determined
in a couple of long running scans.
At three points $(P,T)$ with equal viscosity,
the quasielastic spectra all coincide (Fig.~1a)
which confirms that the microscopic relaxation time is 
proportional to the macroscopic viscosity $\eta$,
and demonstrates that
within experimental error the spectral distribution is independent of pressure. 
We deduce an average slope for isokinetic curves 
$\left( dT/dP \right)_{\tau}$ in the $P,T$ plane of about 
0.28$\pm$0.01\,K/MPa.
Similar values were found for the pressure dependence of the glass transition 
temperature $\left( dT_g/dP \right)$ \cite{Atake_79}, and of the
$\alpha$-relaxation time 
in photon correlation \cite{Fytas_83} 
and dielectric spectroscopy \cite{Naoki_87}.

The invariance of the spectral shape is further illustrated by 
construction of a master curve.
The intermediate scattering function $S(Q,t)$ is obtained by Fourier 
deconvolution of the measured $S(Q,\omega)$.
Since the $\alpha$-relaxation time is proportional to viscosity,
reduced times can be defined via ${\hat t}=t \eta(P_0)/{\eta(P)}$.
The rescaled $S(Q,\hat t)$ coincide over two decades (Fig.~1b).

Time-temperature-pressure superposition is well established by
photon correlation and dielectric loss spectroscopy
in OTP \cite{Fytas_83,Naoki_87} 
as well as in other fragile liquids \cite{otherLiquids};
for a given time shift,
the spectral lineshape changes even less with pressure
than with temperature \cite{Fytas_83,Naoki_87}.
Our measurements extend the validity of this scaling principle
to a picosecond range where contact can be made with fast molecular motion.


Via the total intensity of the elastic or quasielastic
line, backscattering may reveal the presence of
faster relaxation processes.
In fact, the first manifestation of a dynamic transition at $T_c$ 
was obtained from elastic scans \cite{BartschPetry}.
Here, we monitored the pressure dependence of the elastic scattering
$S(Q,\omega\simeq0)$ at 301\,K (Fig.~2).
At high pressures, in the glassy phase, the elastic intensity is constant.
This make the analysis even simpler than in temperature-dependent measurements
where the harmonic evolution of the Debye-Waller factor 
had to be taken into account.
On releasing the pressure, as on increasing the temperature, an anomalous
decrease of the elastic intensity indicates the onset of $\beta$ relaxation.
This decrease is observed first and most pronounced for large wavenumbers.
For the lowest pressures the elastic line begins to broaden, and its total
intensity can no longer be determined from $S(Q,\omega\simeq0)$.
Instead, one has to take the integral over the full $\alpha$ line.

{}From the scaling analysis of isothermal $\alpha$-relaxation 
spectra at 320\,K (Fig.~1b),  
we find that up to 105\,MPa
the integrated $\alpha$-relaxation amplitude $f_Q$ does not depend on pressure.
This allows us to extrapolate $f_Q$ 
at 301\,K from $P_0$ towards higher pressures,
depicted as horizontal lines in Fig.~\ref{Fig2}.
The intersection with the elastic intensities (vertical line) 
which turns out to be independent of~$Q$ 
permits the identification of a critical pressure $P_c\simeq42$ MPa.


On IN6 we access the $\beta$ process directly.
As the quasielastic broadening is of the same order as the instrument's
resolution width, data are analysed by Fourier deconvolution (Fig.~3). 
In our temperature-dependent experiments,
spectra from different instruments were combined into
an intermediate scattering function $S(Q,t)$
which covered three decades in time \cite{WuSqt,ToSqt}; 
the fast $\beta$ relaxation could be described 
by the mode-coupling prediction \cite{Gotze_90}
\begin{equation}\label{scaling_law}
   S(Q,t)=f_Q+H_Q g_{\lambda}(t/t_{\sigma}).
\end{equation}
The dynamic range of the present study does not allow 
an independent verification of (\ref{scaling_law});
nevertheless,
taking advantage of previous results,
we can use (\ref{scaling_law}) to extract the pressure dependence
of~$H_Q$ and~$t_{\sigma}$ from IN6 data.
Besides the shape parameter $\lambda=0.77$
we input the Debye-Waller factor
determined from a Kohlrausch fit to the long-time tail 
of low-pressure data.
While the asymptotic law (\ref{scaling_law}) does not cover the 
short-time dynamics below 1 psec,
it consistently describes the bending of $S(Q,t)$ 
into and out of the intermediate plateau~$f_Q$ (Fig.~3).

Within mode-coupling theory,
the $\alpha$-relaxation amplitude $f_Q$, 
the $\beta$-relaxation amplitude $H_Q$, 
and the cross-over time $t_\sigma$
depend on the equilibrium structure factor~$S(Q)$
which in turn is a regular function of para\-meters like $T$ and~$P$.
Close to the dynamic cross-over,
this dependence can be expressed 
through a linear separation parameter \cite{Gotze_85},
$\sigma\propto(T-T_c)/T_c$ in isochoric 
or $\sigma\propto(P_c-P)/P_c$ in isothermal experiments.
On approaching the cross-over from the liquid side,
$f_Q$ varies only weakly, whereas $H_Q$ and $t_\sigma$ 
are predicted to become singular:
\begin{equation}\label{hq}
  H_Q \propto {\sigma}^{1/2} \quad 
    \mbox{and} 
  \quad t_{\sigma} \propto \sigma^{-1/2a},
\end{equation}
with a critical exponent $a=0.295$ determined by $\lambda$. 

In Fig.~4a,b these power laws are tested for two isotherms
by plotting ${H_Q}^2$ and $t_{\sigma}^{-2a}$ vs.\ $P$.
Asymptotically, linear behaviour is observed,
and extrapolation gives consistent estimates for $P_c$ at 316 and 324\,K.
Away from the asymptotic limit, the data scatter considerably,
similarly as in temperature-dependent experiments \cite{WuSqt,ToSqt}.

Together with $T_c$ at ambient pressure and the value $P_c(301\,{\rm K})$
obtained from elastic scans,
a total of four points of the dynamic phase boundary $T_c(P)$ is obtained.
These points, summarised in Tab.~1,
fall onto a straight line with slope $dT_c/dP\simeq0.28$\,K/MPa
which is in remarkable accord with the slope determined
from $\alpha$-relaxation, viscosity and $T_g$ data:
the mode-coupling cross-over line is parallel to 
lines of equal $\alpha$ response.

Lines of constant density, 
on the other hand, have slopes of about $0.70$\,K/MPa
which confirms \cite{ToSW97a}
that temperature influences the dynamics not only via free volume.
The same conclusion was reached by
simulation of a Lennard-Jones system \cite{Beng86a} and 
by light scattering in cumene \cite{Li_95};
it was suggested instead that at least the $\alpha$ dynamics can 
be parametrised by viscosity \cite{Li_95}.

Our results imply that 
the dynamic cross-over can still be characterised by
one single separation parameter $\sigma(T,P)$ 
which however does not depend on density alone.
Instead, it appears that 
density~$n$ and temperature~$T$ can be combined to
an effective coupling $\Gamma \propto n T^{-1/4}$:
Tab.~1 shows that $\Gamma_c$ is constant within experimental error.
Furthermore, 
constraint fits to $S(Q,t)$ confirm
that the temperature and pressure dependence of amplitude and time scale 
can be described by the power laws (\ref{hq}) 
in conjunction with just one separation parameter
$\sigma = (\Gamma-\Gamma_c)/\Gamma_c$.

A coefficient~$\Gamma\propto n T^{-1/4}$ 
is known to characterise equilibrium properties
of a dense soft-sphere fluid with
repulsive core potentials of the Lennard-Jones $r^{-12}$ type \cite{HaMD86}.
In a supercooled liquid,
we are actually in a high-density regime
where the static structure factor 
$S(Q)$ is sensitive only to the repulsive part of the potential 
\cite{HaMD86,BoYi80}.
In simulations of binary soft-sphere
\cite{WeicheKugeln} and Lennard-Jones \cite{Nauroth_97} systems,
the glass transition was found to occur at constant~$\Gamma_c$.

It was not expected 
that these results apply literally in a complex
molecular liquid like OTP.
In fact, the comparison with a Lennard-Jones liquid 
can be made almost quantitative. 
To this end, we need the two parameters of the $6-12$ potential:
a van-der-Waals diameter $\sigma=7.6$\,\AA,
estimated from structural information \cite{Bond64},
is in good accord with the recently observed prepeak 
in the static structure factor at 
$Q \simeq 0.85\ldots0.9$\,\AA$^{-1}$ \cite{ToSqt,Masciove97},
and a potential depth $\epsilon=600$\,K has been determined
by calibrating a three-site OTP model to experimental density and diffusivity
\cite{LeWa}.
Using this input,
we obtain $\Gamma_c=1.50$ (Tab.~1)
which is closer to the Lennard-Jones result 
$\Gamma^{\rm LJ}_c\simeq 1.25$ \cite{Beng86a,Nauroth_97}
than one could reasonably expect.


In conclusion, our experiments 
give a consistent picture of microscopic dynamics in the $T,P$-plane.

Time-temperature-pressure superposition,
known from Hz--MHz spectroscopy, 
describes structural $\alpha$ relaxation down to the picosecond range
where macroscopic properties of the material 
evolve from its microscopic dynamics.
Spectral lineshapes observed by incoherent neutron scattering
are invariant, 
the mean relaxation time is proportional to the viscosity,
and states of equal kinetic are connected by the same slope $dT/dP$,
regardless of the time scale.

Fast $\beta$ relaxation is detected from the
$\alpha$-relaxation amplitude; 
its full spectral shape is observed by time-of-flight spectroscopy.
Just as in variable temperature measurements,
the data are fully compatible with mode-coupling predictions;
they allow the unambigous and consistent determination of a dynamic cross-over 
line $T_c(P)$.
This line is parallel to the lines of equal viscosity:
$\alpha$ and $\beta$ relaxation are driven by the same parameter.

At least in a certain neighbourhood of $T_c(P)$
this parameter is linear in the effective coupling $\Gamma\propto n T^{-1/4}$ 
which in turn depends ultimately on the molecules' short-range interaction.


We gratefully acknowledge stimulating discussions with
C.~Alba-Simionesco, G.~Diezemann, W.~Kob and W.~Petry. 
We thank A.~Doerk for help during sample preparation, 
K.~Schug for communicating to us results prior to publication,
and W.~G\"otze and W.~Kob for valuable comments on the manuscript.
Special thanks to L.~Melesi for the great support with the pressure equipment.
Financial support by German BMBF 
under 03{\sc fu}4{\sc dor}4 and 03{\sc pe}4{\sc tum}9
is appreciated.



\end{multicols}
\newpage

\begin{table}
\begin{tabular}[th]{|c|c|c|c|}
\hline
$T\;$(K) & $P_c\;$(MPa) & $n_c\;$(nm$^{-3}$) & 
$\Gamma_c$\\
\hline
\hline
290 & ~~0.1  & 2.840 & 1.495(4) \\ 
301 & ~42(3) & 2.875 & 1.499(6) \\ 
316 & ~90(7) & 2.906 & 1.497(5) \\ 
324 & 123(7) & 2.931 & 1.501(9) \\ 
\hline
\end{tabular}

\caption{Critical pressure $P_c$ and corresponding density $n_c$
\protect\cite{n_von_P}
as function of temperature.
The product $\Gamma_c= n \sigma^3 {(\epsilon/T)}^{1/4}$ 
is found to be constant (its errors are estimated  
from $\Delta T_c=\pm3$K at 290 K and $\Delta P_c=\pm5$ MPa).
The choice of $\sigma=7.6\,$\AA\ and $\epsilon=600\,$K
is explained in the text.}
\label{Gamma}
\end{table}

\begin{figure}
\caption{(a) Spectra from IN16 at $Q$=1.8\,\AA$^{-1}$ for three 
         combinations of temperature and pressure which lead approximately 
         to the same viscosity.
         The spectra are scaled to their value at $\omega$=0.
         The line represents the measured resolution function. ---
         (b) Master curve $S(Q,\hat t)$ from isothermal (320\,K) data
         constructed by rescaling times with viscosity.
         The solid line is a fit with the Kohlrausch function 
         $f_Q \exp(-{\hat t}^\beta)$,
         with a stretching exponent~$\beta=0.6$.}
\label{Fig1}
\end{figure}

\begin{figure}
\caption{Debye-Waller factor as a function of pressure at 301\,K.
        Symbols show elastic scattering intensities:
        as long as there is no quasielastic broadening,
        they yield the Debye-Waller factor
        which is $P$ independent in the glass (solid lines).
        Dashed horizontal lines indicate the integrated $\alpha$ intensity,
        measured at $P_0$ and shown to be $P$ independent.
        The dotted vertical line indicates the critical pressure~$P_c$.}
\label{Fig2}
\end{figure}

\begin{figure}
\caption{Intermediate scattering function $S(Q,t)$ at 
        $Q$=1.8\,\AA$^{-1}$ from IN6 in the $\beta$-relaxation regime
        for different pressures.
        Lines are fits with 
        (\protect\ref{scaling_law}) with a fixed shape parameter $\lambda=0.77$
        and fixed plateau values~$f_Q$ (indicated by thick circles).}
\label{Fig3}
\end{figure}

\begin{figure}
\caption{$\beta$-relaxation amplitude and cross-over time 
        for two isotherms 316 and 324\,K and three wave numbers 
        $Q$=1.2 ($\Box$), 1.5 ($\bullet$), and 1.8\,\AA$^{-1}$ ($\diamond$).
        The data are linearised as $H_{Q}^2$ and $t_{\sigma}^{-2a}$ 
        according to (\protect\ref{hq});
        lines are fits to $Q$~averaged values.}
\label{Fig4}
\end{figure}


\begin{thebibliography}{99}
\bibitem{eMCT}
Throughout this letter, we refer to lowest-order 
(idealised) mode-coupling theory.
Within extended theory, a lower value $T_c=276\,$K is found 
\protect\cite{Cumm}.
\end{thebibliography}
\end{document}